\documentclass{nature}

\usepackage{tabulary,graphicx,amsmath,amsfonts,amssymb}
\usepackage[utf8x]{inputenc}
\usepackage{mathtools}
\usepackage{color}
\usepackage{subcaption}
\definecolor{blueryb}{rgb}{0.01, 0.28, 1.0}

\definecolor{greenryb}{rgb}{0.1, 1.0, 0.1}

\makeatletter
\renewenvironment{figure}
               {\@float{figure}}
               {\end@float}
\renewenvironment{figure*}
               {\@dblfloat{figure}}
               {\end@dblfloat}
\renewenvironment{table}
               {\@float{table}}
               {\end@float}
\renewenvironment{table*}
               {\@dblfloat{table}}
               {\end@dblfloat}

\makeatother

\usepackage{url,multirow,morefloats,floatflt,cancel}
\makeatletter

\AtBeginDocument{\@ifpackageloaded{textcomp}{}{\usepackage{textcomp}}}
\makeatother
\usepackage{colortbl}
\usepackage{xcolor}
\usepackage{pifont}
\usepackage[nointegrals]{wasysym}
\makeatletter

\def\mcWidth#1{\csname TY@F#1\endcsname+\tabcolsep}

\def\cAlignHack{\rightskip\@flushglue\leftskip\@flushglue\parindent\z@\parfillskip\z@skip}
\def\rAlignHack{\rightskip\z@skip\leftskip\@flushglue \parindent\z@\parfillskip\z@skip}

\if@twocolumn\usepackage{dblfloatfix}\fi 
\AtBeginDocument{
\expandafter\ifx\csname eqalign\endcsname\relax
\def\eqalign#1{\null\vcenter{\def\\{\cr}\openup\jot\m@th
  \ialign{\strut$\displaystyle{##}$\hfil&$\displaystyle{{}##}$\hfil
      \crcr#1\crcr}}\,}
\fi
}

\AtBeginDocument{%
  \@ifpackageloaded{endfloat}%
   {\renewcommand\efloat@iwrite[1]{\immediate\expandafter\protected@write\csname efloat@post#1\endcsname{}}}{}%
}%

\let\lt=<
\let\gt=>
\def\processVert{\ifmmode|\else\textbar\fi}

\@ifundefined{subparagraph}{
\def\subparagraph{\@startsection{paragraph}{5}{2\parindent}{0ex plus 0.1ex minus 0.1ex}%
{0ex}{\normalfont\small\itshape}}%
}{}

\newcommand\role[1]{\unskip}
\newcommand\aucollab[1]{\unskip}
  
\@ifundefined{tsGraphicsScaleX}{\gdef\tsGraphicsScaleX{1}}{}
\@ifundefined{tsGraphicsScaleY}{\gdef\tsGraphicsScaleY{.9}}{}
\def\checkGraphicsWidth{\ifdim\Gin@nat@width>\linewidth
	\tsGraphicsScaleX\linewidth\else\Gin@nat@width\fi}

\def\checkGraphicsHeight{\ifdim\Gin@nat@height>.9\textheight
	\tsGraphicsScaleY\textheight\else\Gin@nat@height\fi}

\def\fixFloatSize#1{}
\let\ts@includegraphics\includegraphics

\def\inlinegraphic[#1]#2{{\edef\@tempa{#1}\edef\baseline@shift{\ifx\@tempa\@empty0\else#1\fi}\edef\tempZ{\the\numexpr(\numexpr(\baseline@shift*\f@size/100))}\protect\raisebox{\tempZ pt}{\ts@includegraphics{#2}}}}

\AtBeginDocument{\def\includegraphics{\@ifnextchar[{\ts@includegraphics}{\ts@includegraphics[width=\checkGraphicsWidth,height=\checkGraphicsHeight,keepaspectratio]}}}

\def\URL#1#2{\@ifundefined{href}{#2}{\href{#1}{#2}}}

\def\UrlOrds{\do\*\do\-\do\~\do\'\do\"\do\-}%
\g@addto@macro{\UrlBreaks}{\UrlOrds}
\makeatother

\def\fixFloatSize#1{}
\begin{document}

\title{Understanding high pressure hydrogen with a hierarchical machine-learned potential}
  
\author{Hongxiang Zong$^{1,2}$, Heather Wiebe$^1$ and Graeme J. Ackland$^1$,}

\maketitle 

\begin{affiliations}
\item Centre for Science at Extreme Conditions and School of Physics and Astronomy,  University of Edinburgh, Edinburgh, EH9 3ET, UK
\item State Key Laboratory for Mechanical Behavior of Materials, Xi’an  
Jiaotong University, Xi’an, Shanxi 710049, China
\end{affiliations}


\begin{abstract}
The hydrogen phase diagram has a number of unusual features which are generally well reproduced by density functional calculations.  Unfortunately, these calculations fail to provide good physical insights 
into why those features occur.  In this paper, we parameterize a model potential for molecular hydrogen which permits long and large simulations.  The model shows excellent reproduction of the phase diagram, including the broken-symmetry Phase II, an efficiently-packed phase III and the maximum in the melt curve.  It also gives an excellent reproduction of the vibrational frequencies, including the maximum in the vibrational frequency $\nu(P)$ and negative thermal expansion.   By detailed study of lengthy molecular dynamics, we give intuitive explanations for observed and calculated properties. All solid structures approximate to hexagonal close packed, with symmetry broken by molecular orientation. At high pressure, Phase I shows significant short-ranged correlations between molecular orientations.
The turnover in Raman frequency is due to increased coupling between neighboring molecules, rather than weakening of the bond. The liquid is denser than the close-packed solid because, at molecular separations below 2.3\AA, the favoured relative orientation switches from quadrupole-energy-minimising  to  steric-repulsion-minimising.  The latter allows molecules to get closer together, without atoms getting closer but this cannot be achieved within the constraints of a close-packed layer.
\end{abstract}

\clearpage
\section{Introduction}
Since the discovery of solid molecular hydrogen in 1899, the nature of this phase has remained controversial\cite{dewar1899solidification}.   It is now believed that the solid "Phase I" comprises rotating hydrogen molecules on a hexagonal close-packed lattice\cite{VanKranendonk1983}.  With increasing pressure the rotation becomes hindered\cite{lorenzana1990orientational} by intermolecular interactions, both steric and electrostatic, leading ultimately to phase transformations to a low temperature Phase II\cite{silvera1981new}, in which quadrupole-quadrupole interactions (EQQ) arrest the rotation\cite{van2020quadrupole}, and a high pressure Phase III\cite{hemley1988phase,lorenzana1989evidence}, in which steric interactions dominate.  

Experimental study of these phases has proved challenging.  Most information is gleaned from spectroscopy, with the first room temperature X-ray study only completed in 2019\cite{akahama2010evidence,ji2019ultrahigh}. Raman spectroscopy shows peaks corresponding to quantum rotors at low pressure, which gradually broaden and shift with pressure, and a distinctive sharp phonon mode which rules out cubic close packing as a structure\cite{goncharov1998new,goncharov2001spectroscopic,gregoryanz2003raman,akahama2010raman,zha2014raman,howie2015raman}.
The melt line has a strongly positive Clapeyron slope at low pressures, with a turnover around 100GPa\cite{bonev2004quantum,deemyad2008melting,eremets2009evidence,geng2015melt}. The negative slope means that even though the solid is hexagonal "close-packed", the liquid must be even denser.  The turnover also means the liquid has higher compressibility, but how this comes about remains unexplained.  X-ray studies at low temperature traversing Phase I-II-III do not show any convincing structural changes, in part because it has proven impossible to get sufficient resolution to determine the molecular orientation\cite{akahama2010evidence,ji2019ultrahigh}.   

Spectroscopy gives vibrational data, which are still insufficient to determine the structures of phases II, III and IV. There have been many and varied attempts to identify the structures via simulations\cite{kohanoff1997solid,kohanoff1999dipole,pickard2007structure,pickard2009structures,martinez2009novel,liu2012room,magdau2013identification,monserrat2016hexagonal,magdau2016phaseV,magdau2017simple,magdau2017infrared}. However, a consensus has not yet been reached. Based on fully ab initio calculations, including density functional theory (DFT) or quantum Monte Carlo (QMC)\cite{drummond2015quantum}, a number of candidate structures have been proposed for phase II and III. Besides differing molecular orientation, they are all similar, consisting of  primitive cells with lattice sites close to hcp\cite{pickard2007structure}. Among the structures, the P21/c-24, C2/c-24, and Pc-48 structures provide low-energy candidate structures for phases II, III, and IV. 

The modern theory of the structure of these phases is based around electronic structure calculations.  The early work involved calculating the ground state, assuming classical nuclei, then adding quantum-nuclear effects via the quasiharmonic approximation.  This methodology, whether based on DFT or QMC, predicts hcp-like ground states for Phases I-III in agreement with X-ray data.  However the spectroscopic signature of the Phase II - the appearance of many sharp, low-frequency, peaks\cite{liu2017high,goncharov2001spectroscopic} - is not well reproduced by the quasiharmonic calculations.  As explained in the previous paragraph, the likely cause is a failure of the harmonic mode assumption for excited states, rather than the DFT itself.   

To understand the high-temperature phases, one needs to examine non-harmonic behaviour, including rotation, which means going beyond a single unit cell, e.g. using molecular dynamics. Molecular dynamics requires forces on each atom based on the positions of all the atoms in the system, which requires a force model which is fast enough to allow large simulations. Here we use a machine-learning approach to derive a transferable force model based on an interatomic potential.  
There are several approaches to machine learning interatomic forces\cite{behler2007generalized,bartok2010gaussian,botu2015adaptive}, which balance speed, transferability and accuracy.  We adopt an approach focusing on transferability.  

The machine-learned potential should conserve energy, and therefore be based on a Hamiltonian (the potential). 
Forces are guaranteed to be conservative if they depend on translational and rotational invariant quantities: the "fingerprint" of each atom.  We are interested in molecular phases here, so our potential specifies which atoms are "bonded" and allows stretching but not breaking of bonds. 

For hydrogen the machine-learning approach is trained on energies and classical (Hellmann-Feynman) forces derived from standard density functional theory. In the Born-Oppenheimer approximation adopted by all standard DFT codes, the interatomic potential is the same for deuterium, HD and hydrogen. Contributions from quantum-nuclear effects can be incorporated using lattice dynamics or path integral methods.

\section{Results}

\subsection{Fitting Forces and the Phase diagram:}

A particular challenge for hydrogen comes from the hierarchy of energies. The covalent bond is much stronger than the van der Waals attraction between molecules, which is turn is much stronger than the EQQ interactions which determine molecular orientations.   To address this our potential combines a hierarchical fitting strategy alongside machine learning (HMLP)  described below. 

For transferability testing, we used the standard approach of fitting to a subset of the data and testing against a different subset.  Furthermore, we used an iterative fitting process: a trial potential was fitted, and applied in both Phase II annealing and melting line MD simulations. If novel configurations were found, they were used to generate more reference states for the DFT database, and the fitting process was repeated.  This iterative process ensures that spurious structures are suppressed and the ground state structure is the same as found in DFT. 

Technical details of the forcefield parameterization are given in the methods section. 

\subsection{Phase diagram:}
Fig. \ref{fig:Phase-Diagram} shows the very good agreement between the classical HMLP and the DFT phase diagrams.  By eliminating finite size effects, the HMLP can capture the full long-range correlations, however, this does not appear to have a significant effect on the phase boundaries.

In Phase I, the H$_2$ molecules keep free rotation at pressures below 40 GPa and temperatures below 900 K (orange hexagon symbols), and at higher pressures the rotation is inhibited but there is no long-ranged order. At low temperatures, phase II becomes stable (red triangles), and the stable temperature region increases gradually with pressures. At high temperatures, the hcp lattice collapses to a liquid state (blue squares). The calculated melting curve has a strong positive slope ($dT/dP >0$) at low pressures, reaches a maximum at around 900 K and 90 GPa, and then drops.  The HMLP predicted phase diagram agrees reasonably with experimental observations, as well as DFT (Fig. \ref{fig:Phase-Diagram}). 

The HMLP and DFT predictions are good for the melt curve, but both overstabilise the broken-symmetry Phase II. This is due to the lack of quantum nuclear effects, notably the zero-point energy, and can be addressed by including quantum nuclear effects in the simulation.  The discrepancy in the Phase I-II line does not indicate any inaccuracy of the HMLP itself.
 Our predicted melting curve is consistent with experiments: the value for the melting curve maximum is located between 80-100 GPa and 900 K, similar to the HMLP potential values. It also agrees with two-phase ab initio simulations, which proposed a gradual softening of the intermolecular repulsive interactions as its cause \cite{bonev2004quantum}. The close agreement of the HMLP transition pressure with experimental data enables us to accurately simulate behaviors of temperature- or pressure- driven phase transition between phase I and II. A "Phase III" is observed at higher pressures, corresponding to a different symmetry-breaking.  However, by design the present HMLP model should start to fail to capture the properties of H$_2$ at still higher pressures, where molecule dissociation needs to be considered. 

\subsection{Nature of Phase I}

Phase I can be easily recognised in MD by ordering of the molecular centres on the hcp lattice, and disorder of the orientations.  Although frequently referred to as a free rotor phase, we find this to be true only at low pressures. As pressure is increased the angular momentum autocorrelation becomes shorter than a single rotation, and then acquires a negative component, indicating that the molecule is librating.

 Another characteristic of Phase I is the molecular vibration or "vibron": in Raman scattering this corresponds to the in-phase vibration of all molecules. The vibron frequency first increases, then decreases with pressure.
 Two plausible reasons are given for this reduction: either increased intermolecular coupling or weakening of the covalent bond. In our model, the covalent bond is always described by the same Morse parameters, so changes in the vibron frequency can arise only from resonant interactions between molecules, not weakening of the bond.  Thus reproducing the reentrant vibron behaviour is a test of both the physical basis and the parameterization of the model.

Since the molecules are rotating in Phase I, lattice dynamics cannot be used, so Raman phonon frequency is numerically characterised by the in-phase mode-projected velocity auto-correlation function (VAF) \cite{ackland2014efficacious}. Trajectories and velocities were produced from 150 K HMLP-MD simulations within the micro-canonical ensemble (NVT) initiated in the P21/c structure. A very fine time step of 0.05 fs was used and the trajectory and velocities were saved every 10 time steps.  Using the projection method and Fourier transformation of the VAFs we can calculate both the total vibron density of states from
\begin{equation}
    g_{tot}(\omega) = \sum_{ik} \int [r^2_{ik}(t)] \exp{i\omega t}
\end{equation}
and the signal from the most strongly Raman-active mode,
\begin{equation}
   g_{Raman}(\omega) = \int \left [\sum_{ik} r^2_{ik}(t) \right ] \exp{i\omega t} 
\end{equation}
where $ik$ runs over all molecules (comprising atoms $i$ and $k$).  A similar projection method is used for the E2g phonon\cite{pinsook1999calculation}.

\begin{figure}[ht]
\centering
\includegraphics[width=0.8\columnwidth]{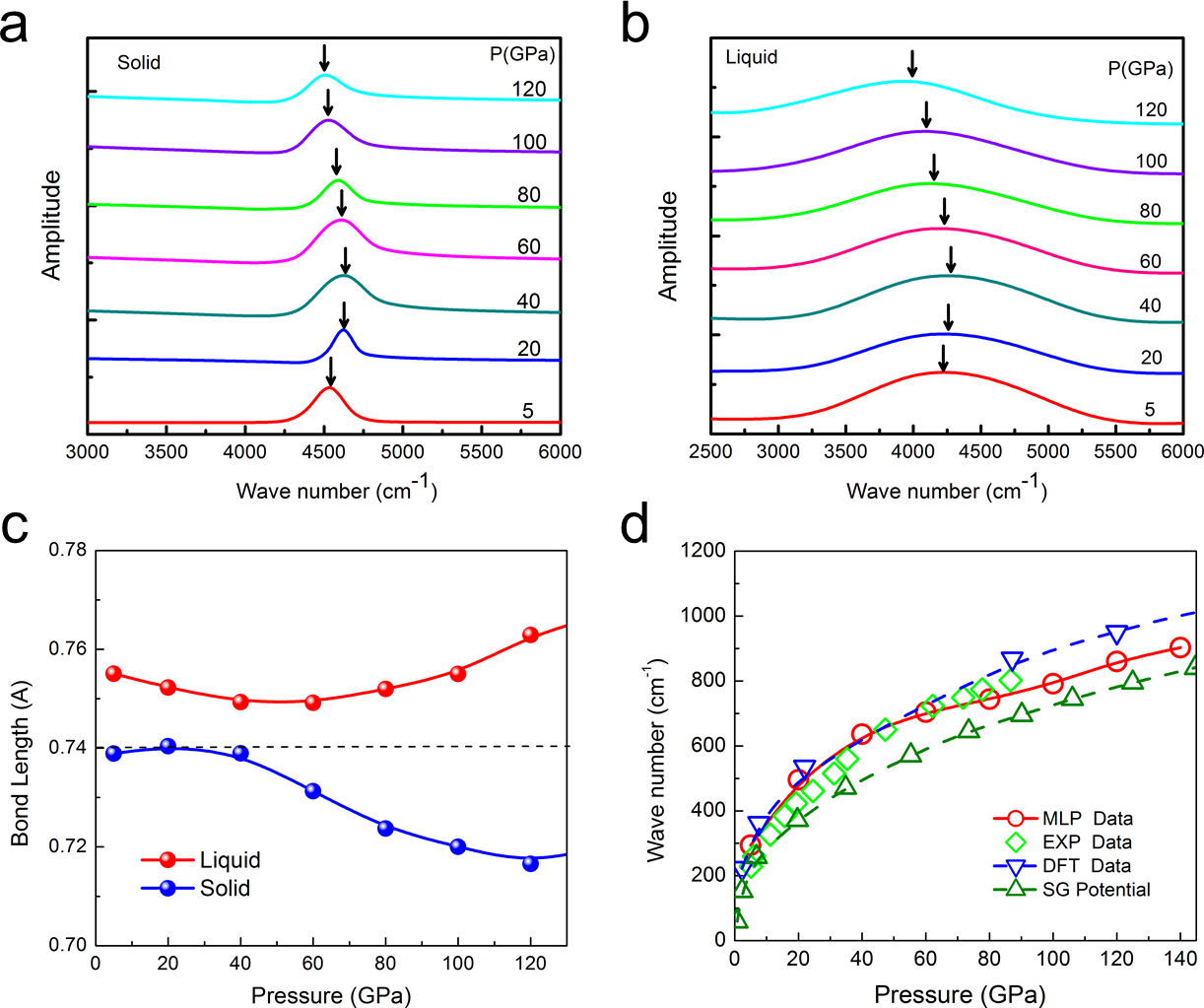}
\caption{Calculated Raman signals of solid and liquid H$_2$ as a function of pressures. \textbf{a} Pressure dependence of vibron frequencies from the solid H$_2$ at T= 150 K, arrows emphasise the turnover of the peak frequency; \textbf{b} Pressure dependence of vibron frequencies from the liquid H$_2$ at T = 1000 K; \textbf{c} The mean H-H bond length as a function of pressures; \textbf{d} The pressure dependent E2g phonon frequency compared with our DFT, recent (to be published) experiments by Pena-Alvarez, and Silvera-Goldman potential\cite{silvera1978isotropic,wijngaarden1983pressure}}
\label{fig:Raman_signal}
\end{figure}

Fig. \ref{fig:Raman_signal} plots the calculated total vibron spectra of solid and liquid hydrogen as a function of pressure from MD simulations. Both show a signature of vibron turnover  above a critical pressure (about 54 GPa), consistent with the experimental observations\cite{goncharov2001spectroscopic}. 
This proves that bond weakening is not required for the turnover, since our potential has a fixed bond strength. Notably, the mean bondlength in phase I decreases monotonically with pressure (Fig. \ref{fig:Raman_signal}(c) ), again at odds with ideas of bond weakening. What appears to be happening is a competition between two effects: at higher pressures the compression of the bond causes an increase in the frequency due to anharmonicity in the potential, whereas above 54 GPa the frequency is lowered due to coupling between the molecules. 

 The hcp structure has a Raman-active optical mode (E2g symmetry) in the phonon spectrum which corresponds to the out-of-phase shear motions in the basal plane. The  frequency range of this Raman mode is experimentally well-determined and extremely large, from 36 cm$^{-1}$ at zero pressure to 1100 cm$^{-1}$ at 250 GPa \cite{hanfland1993novel,hemley1990low,mao1994ultrahigh,loubeyre2002optical}.
The comparison between theoretical and experimental pressure dependencies $\nu(P)$ of the E2g optical phonon Raman active mode is presented in Fig. \ref{fig:Raman_signal}(d). The red symbols in are our HMLP-MD predictions, consistent with the DFT data of this mode extracted from our calculations.  Comparing the present theoretical results with experiment we see that the HMLP predicted frequency curves agree with experiment slightly better than existing isotropic empirical potentials\cite{silvera1978isotropic,wijngaarden1983pressure} (olive curve).

\subsection{Denser than close-packed liquid}

\begin{figure}[ht]
\centering
\includegraphics[width=0.8\columnwidth]{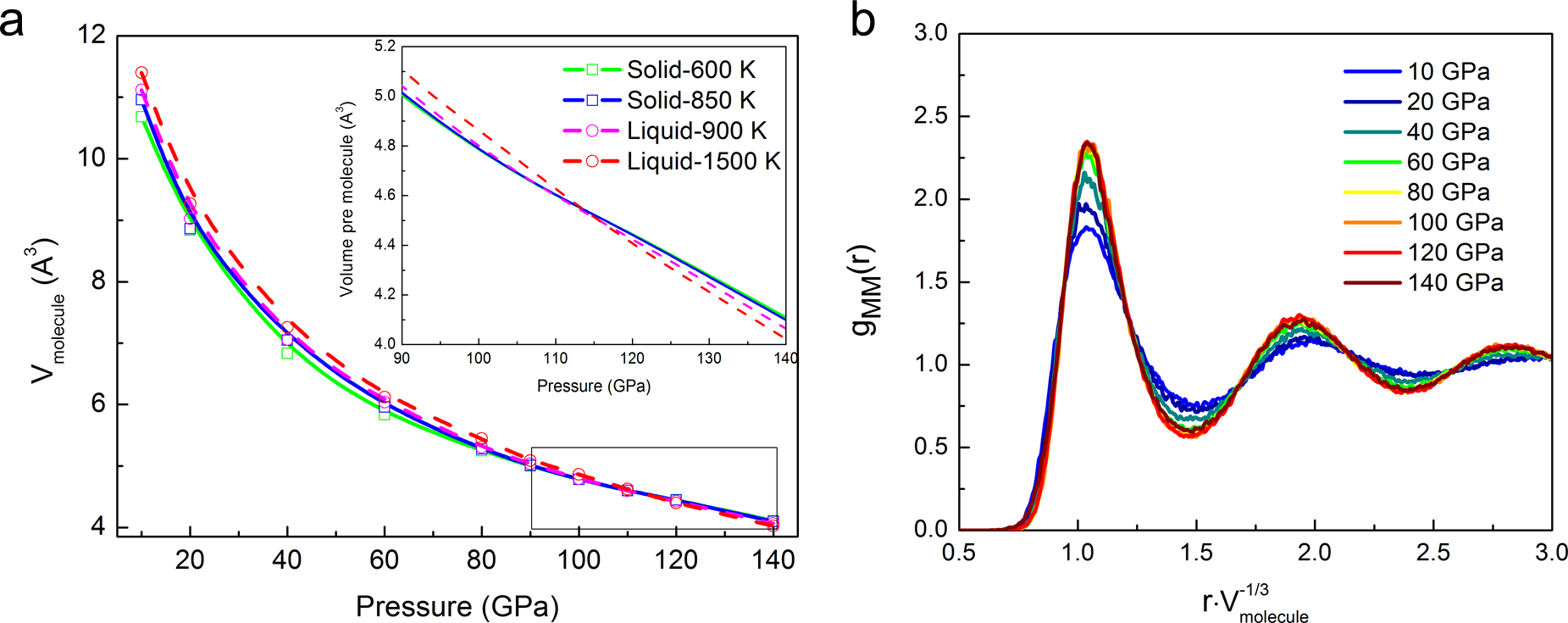}
\caption{Structural properties and EOS of liquid H$_2$ at different densities. \textbf{a} Equation of state (EOS) for liquid and solid H2 at selected temperatures; \textbf{b} Normalised radial distribution
function of molecular centers at selected pressures and T = 1000 K, indicating no liquid-liquid phase transition below 140 GPa.}
\label{fig:LLPT}
\end{figure}

Fig. \ref{fig:Phase-Diagram} shows that the potential correctly reproduces the turnover and negative Clapeyron slope.  We investigated the possible explanation for this denser-than close-packed liquid.  Fig. \ref{fig:LLPT}(a) shows the equation of state for both solid and liquid phases, with the crossover indicating where the liquid is denser than the solid.  The HMLP predicts a negative thermal expansion, which is consistent with DFT\cite{geng2019thermodynamic}.  The normalised radial distribution function  (Fig. \ref{fig:LLPT}(b) ) shows that the liquid structure is essentially unchanged with pressure up to the pressures where bond-breaking becomes a factor. We therefore deduce that the denser liquid is not  related to the molecular-atomic transition.

\begin{figure}[ht]
    \centering
    \includegraphics{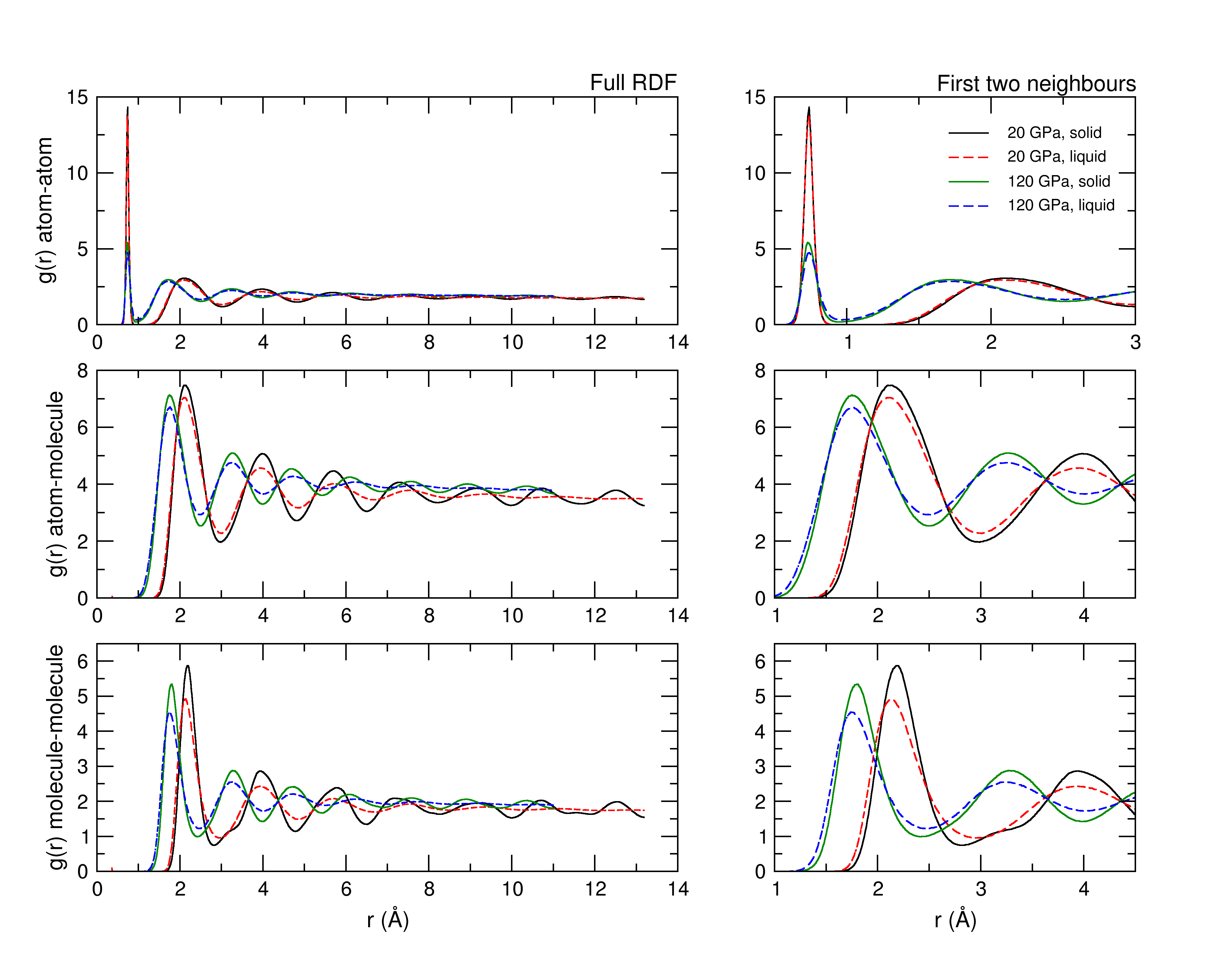}
    \caption{Radial distribution functions (RDFs) of coexisting solid and liquid show that the liquid can pack more tightly than the hcp solid phase. This is true both at 20 GPa, which is before the turnover in the Clapeyron slope, and at 120 GPa, well into the negative slope region.  Shown here are atom-atom, atom-molecule and molecule-molecule RDFs, with the full length shown on the left and a close-up of the first two neighbour shells on the right. The higher density of the liquid is most apparent in the molecule-molecule RDF. These RDFs are on good agreement with the RDFs obtained from AIMD simulations\cite{geng2019thermodynamic}}
    \label{fig:coexistenceRDFs}
\end{figure}

Fig. \ref{fig:coexistenceRDFs} compares the solid and liquid at the melting point. They are remarkably similar: close to the phase boundary the liquid shows five discernable neighbour peaks indicating short ranged structure to 10\AA.  The molecular bondlength is longer in the liquid than the solid (shown more clearly in Fig. \ref{fig:Raman_signal}), but the separation between molecules is noticeably smaller in the liquid as evidenced by the first peak in the molecule-molecule RDF. This means that the molecules get closer together in the liquid, despite being longer.  

 Intermolecular interactions are dominated by quadrupole-quadrupole interactions and steric repulsion. Table \ref{table:EQQ} shows the implied contribution from quadrupole-quadrupole interactions calculated by electrostatics from HMLP sumulations. Althought the ML potential has no explicit electrostatic terms, there is a strong orientation correlation, which lowers the quadrupolar energy, not only in Phase II, but also in Phase I and in the liquid.
 
\begin{figure}[ht]
    \centering
    \includegraphics[width=0.48\columnwidth]{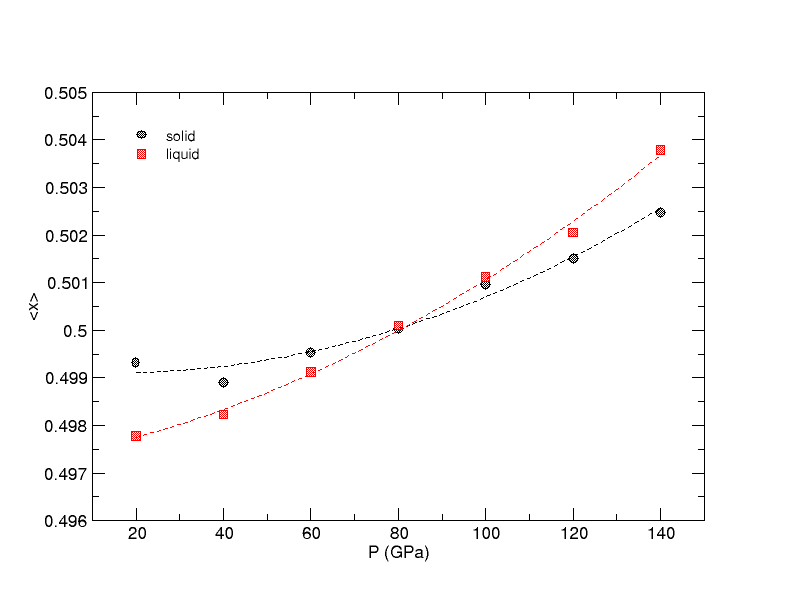}
\includegraphics[width=0.48\columnwidth]{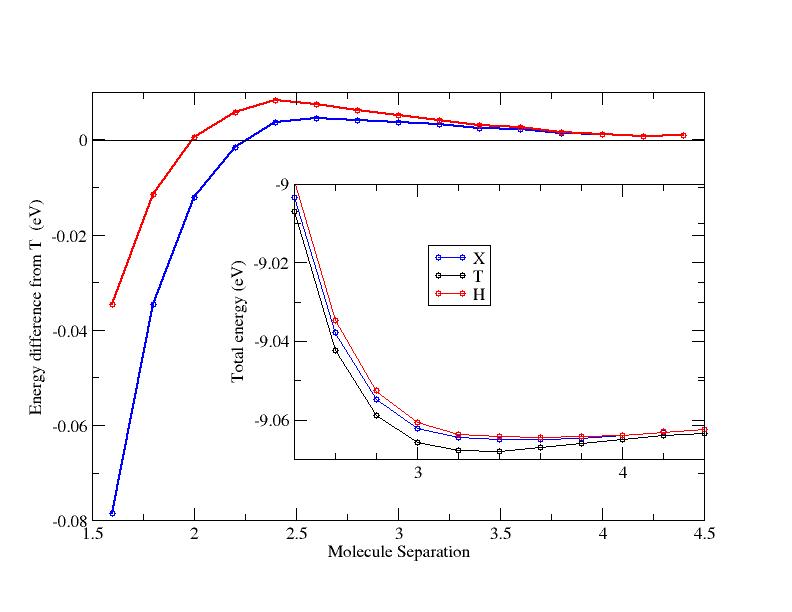}    \caption{(Left) 
A measure of how close to the X configuration the 12 nearest neighbours are (See Methods).
Data was obtained as a time average over solid-liquid coexistence trajectories (500,000 steps, 27648 atoms). 
(Right) DFT calculations of the interaction between two hydrogen molecules as a function of distance for three orientations. Taking the separation vector along the z-axis, and one of the molecules to point along $x$;  H,X, and T  represent the second molecule pointing along $x$, $y$ and $z$ respectively.  The configuration with both pointing along $z$ is always unfavorable. Main figure: energy differences relative to T, Inset: total energy relative to free atoms.}
    \label{fig:EQQdifference}
\end{figure}

 There is little difference in the quadrupole energy between Phase I and liquid, so we investigated an alternative measure of the relative orientation. We hypothesised that molecules can get closer together if their constituent atoms are further apart (i.e. in an "X" shape viewed down the intermolecular vector), and so we calculated the degree of X-character, $<X>$, of the 12 nearest molecular neighbours in both liquid and solid phases at the melt. Fig. \ref{fig:EQQdifference}(a) shows that there is a crossover at 80 GPa, where the liquid becomes more X-like than the solid at high pressures. 
  
 Fig. \ref{fig:EQQdifference}(b) investigates this further using DFT, showing that the  T configuration which optimises the quadrupole interaction becomes unfavourable with respect to the X configuration at a separation of 2.25 \AA. As we saw in Fig. \ref{fig:coexistenceRDFs}, the nearest neighbours are already this close by 20 GPa. 

These findings explain why the liquid is denser than the solid.  The X orientation, favoured at short range, cannot be achieved on the hcp lattice without significant frustration. For a given intermolecular separation, the X configuration maximises the atom-atom distances - this and similar arrangements compensates for the smaller molecule-molecule distance in the liquid to give the same peak position in the atom-atom RDF in both liquid and solid.

\subsection{Nature of  Phase II}

\begin{figure}[ht]
\centering
\includegraphics[width=0.8\columnwidth]{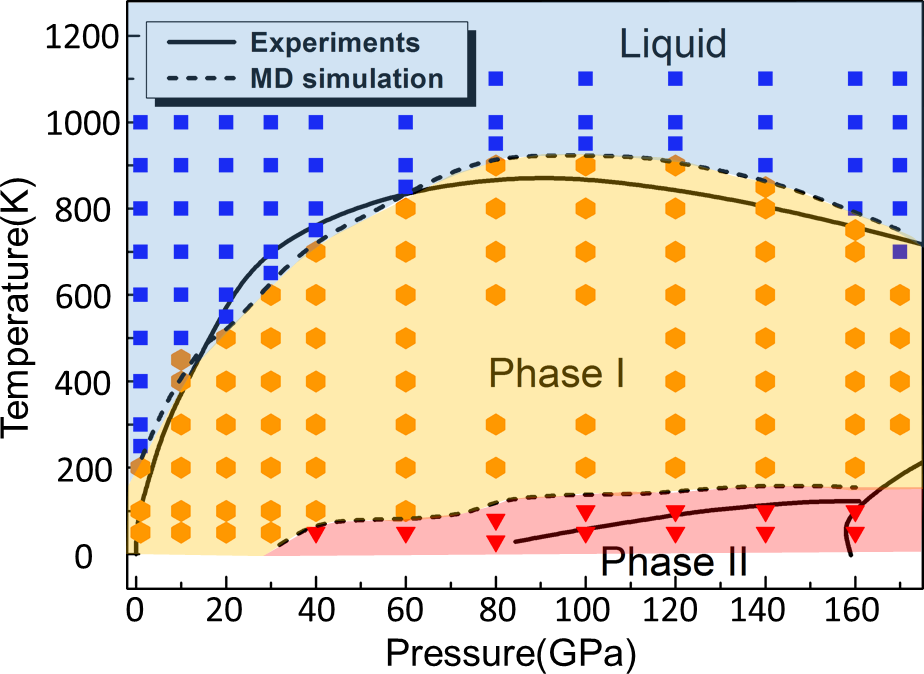}
\caption{Forcefield simulated phase diagram of hydrogen.
Each datapoint represents an HMLP-MD simulation which is itself in  agreement with the equivalent DFT simulation. The dashed  melt line is to guide the eye.
Experimental phase boundaries are taken from Ref.\cite{}.}
\label{fig:Phase-Diagram}
\end{figure}

We performed extensive HMLP-MD simulations around the Phase I-II boundary, with different starting configurations, to determine candidate structures and phase stability of H$_2$ Phase II. At 150 K and 20 GPa, the orientations of the H$_2$ molecular axes are almost randomly distributed along different directions, indicating Phase I with freely rotating molecules.  Upon compression to the high pressure of 80 GPa and cooling to a low temperature of 50 K, the material transforms to an orientationally ordered phase in which the molecular rotations are restricted (Fig. \ref{fig:Phase-Diagram}). By carefully comparing it with candidate structures proposed by ab initio calculations, we find that the lattice and molecular ordering is close to P$2_1/c$, which has been one of the most thoroughly studied and strongest candidate for phase II\cite{pickard2007structure,van2020quadrupole,geng2012high,pickard2009structures}. 

The HMLP does not include quadrupole interactions explicitly - it has "learnt" them.  Table \ref{table:EQQ}  shows what the quadrupole interactions would be, using electrostatic calculation based on the HMLP configurations. The large negative values indicate the prevalence of quadrupole-type ordering: strongest in Phase II. The differences, tens of meV, is of similar magnitude to the phase transformation temperature.  The HMLP has learnt that Phase II is stabilised by quadrupole interactions.

\begin{figure}[!htbp]
    \includegraphics[width=0.8\textwidth]{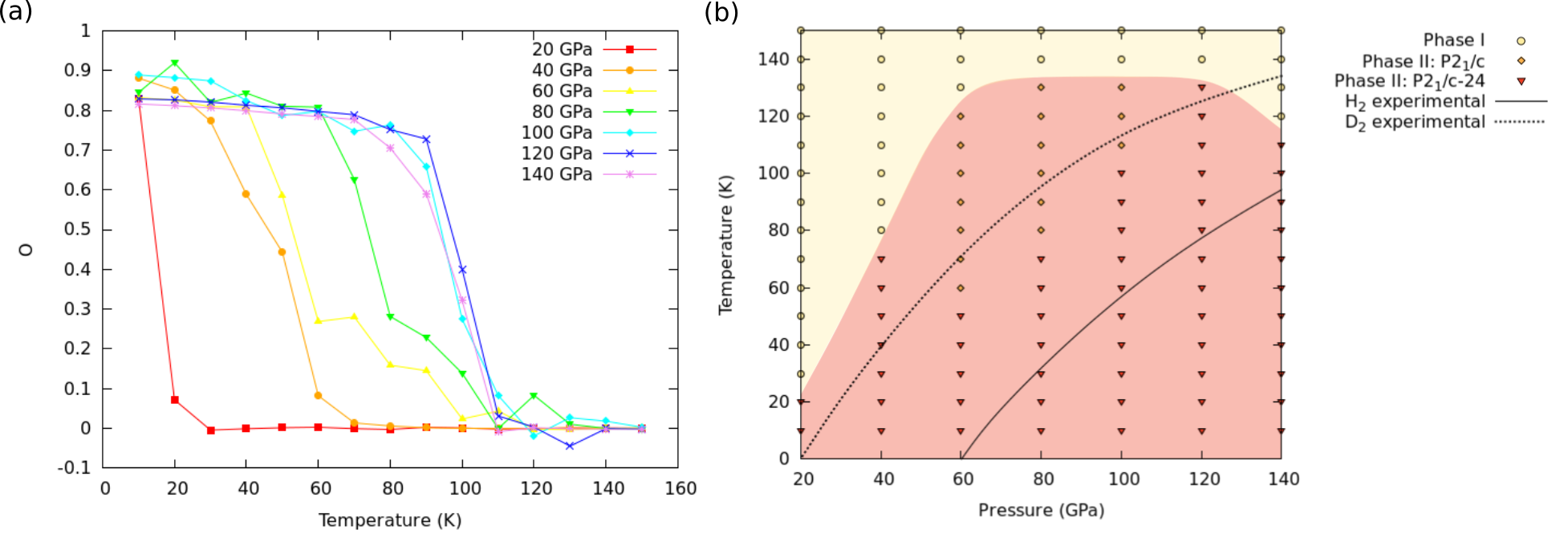}
    \caption{(a) Order parameter $O$ as a function of temperature for the 7 pressures investigated in this work. In all cases there is a sharp decrease from an ordered system ($O=1$) to a disordered system ($O=0$). (b) The resultant phase boundary for the I-II transition in the classical solid. The solid and dashed lines represent the experimental phase boundaries for hydrogen and deuterium respectively\cite{liu2017high}. }
    \label{fig:classicalOP_phaseDiagram}
\end{figure}

The corresponding RDF indicates that the center of each molecule remains close to the hcp lattice sites. Consequently, we define an order parameter $O$ relating the structures of Phase II in our HMLP-MD simulations with the static-lattice DFT predictions of P$2_1/c$-24. $O$ exhibits a sharp change in the order parameter as the system transitions from the structured phase II to the rotationally symmetric phase I with increasing temperature. The intermediate values of $O$ for 60-100 GPa are due to a phase change to a structure with a smaller unit cell (denoted $P2_1/c$ in Fig. \ref{fig:classicalOP_phaseDiagram} to distinguish from $P2_1/c-24$), which for the purposes of locating the phase I-II boundary is still considered to be phase II in Fig. \ref{fig:Phase-Diagram}. Transition temperatures were taken at discontinuous jumps in $O$. This produces the phase diagram shown in Fig. \ref{fig:classicalOP_phaseDiagram}(b). Note that the transition temperatures obtained from analysis of $O$ agree with those obtained from an analysis of the fluctuations of the volume of the system (see Appendix). The phase boundary agrees well with the experiment,\cite{liu2017high} particularly for the more classically-behaved deuterium, and the temperature induced transition within Phase II is similar to the Phase II' identified by Goncharov {\it et al}\cite{goncharov1996raman} in deuterium. Similarity to experiments on deuterium rather than hydrogen is perhaps unsurprising, since nuclear quantum effects (NQE) such as zero point motion are significant at the low temperatures investigated here.

\subsection{Transition to Phase III}

Above 160 GPa we find a high-pressure transformation to a broken-symmetry structure different from Phase II dominated by efficient packing rather than EQQ, this is at approximately the same pressure as Phase III.  

Experimentally, Phase III is associated with a sharp drop in the vibron frequency and the appearance of a strong IR signal. This implies a non-centrosymmetric structure and a weakening of the molecules.  In studies of hydrogen-deuteride (HD) a process of bond dissociation and recombination ({\it "DISREC"}, 2HD $\rightarrow$ H$_2$+D$_2$) has been observed\cite{ dias2016new}.  This bond breaking is seen in DFT to also occur in pure H$_2$\cite{magdau2013identification,magdau2017infrared}.  Our potential does not allow for bond breaking, so we have not studied the dynamics of this "Phase III" in detail. 
 
\subsection{Discussion and Conclusion}

In summary, we have introduced a heirarchical, iterative  machine learning based interatomic potential for atomistic simulations of H$_2$ molecules, by directly learning from reference ab initio molecular dynamics simulations. The resultant HMLP-MD approach predicts angular energy dependence in the range of tens of meV/atom and demonstrates good transferability to various structural environments. Several applications have been presented for which our potential is particularly well suited. 
The fast, transferrable potential is also suitable for a wide range of further applications and extensions, including compounds, bond breaking and path integral calculations.
 
 The simulations reproduce the equilibrium temperature-pressure phase diagram for molecular phases (I, II, III and melt, P $<$ 160 GPa). 
 
 The maximum in the melt curve in hydrogen is highly counterintuitive - it requires that the liquid is denser than the hexagonal-close packed solid. By detailed simulation we resolve this by showing that certain molecular orientations (e.g. X) allow the molecules to approach more closely. The energetically favoured orientational arrangement leads to larger interatomic distances.  With increased temperature, the unfavoured orientations are thermally occupied, leading to a negative thermal expansion in Phase I.  These X-type configurations
 are also incompatible with the hcp lattice, so become more prevalent on melting. This causes the reduced intermolecular distance which gives the densification of the liquid.
 
 Our HMLP potential also has shown the capability of predicting the pressure dependence of the Raman-active E2g mode, consistent with experiment and previous DFT calculations. 
 
We explain the maximum frequency of the vibron as due to competition between molecular compression and stronger intermolecular coupling.  Weakening of the covalent bond is not required.

{}

\begin{table}[ht]
    \caption{\label{table:EQQ} Quadrupole-quadrupole interaction energies $E_{QQ}$ for liquid and solid phases at the melt, alongside values calculated for stable phase I (T = 150 K) and phase II (T = 50 K) structures. All values are in units of meV/molecule. The quadrupole energies of the coexisting solid and liquid phases are considerably lower than either of the stable solid phases, which can be expected due to the high temperature at the melt. At low pressures the solid has a lower quadrupole interaction energy than the liquid, but at $P > 80$ GPa this energy gap starts to decrease and eventually favours the liquid at 140 GPa. For uncorrelated free rotors, $<E_{QQ}^L>$=0 }
    \begin{tabular}{ c c c c c }
         \hline
         \hline
         Pressure (GPa) & $<E_{QQ}^L>_{melt}$ & $<E_{QQ}^S>_{melt}$  & $<E_{QQ}^S>_{Phase I}$ & $<E_{QQ}^S>_{Phase II}$  \\
         \hline
         20 & -6.1845 & -6.2036 & -16.4596 & - \\
         40 & -8.5537 & -8.5077 & -24.8527 & -35.4657 \\
         60 & -10.6602 & -10.7230 & -32.4808 & -42.7504 \\
         80 & -12.2599 & -12.6621 & -41.0739 & -52.1645 \\
         100 & -13.8211 & -14.1031 & -47.2947 & -60.5124 \\
         120 & -15.3890 & -15.3973 & -50.468 & -68.7409 \\
         140 & -17.3734 & -16.8264 & -53.3522 & -76.3981 \\
         \hline
         \hline
    \end{tabular}
\end{table}

\clearpage
\section{Methods}

\subsection{Machine Learning the interatomic potential}

\subsubsection*{Learning data set}
Structures for reference atomic environments and benchmarks were accumulated from density functional theory (DFT)-based ab initio MD runs. The DFT calculations were performed using the CASTEP package\cite{clark2005first} within the Perdew-Burke-Ernzerhof generalised gradient approximation (PBE)\cite{PBE} for the exchange-correlation function. A cutoff energy of 1000 eV for the plane-wave basis set and a k-point mesh of 1x1x1 were selected. To ensure the transferability of the potential to a wide variety of atomistic situations, H$_2$ in different geometric arrangements was considered, including modest-sized bulk samples in phase I, II and liquid, composed of 144 H$_2$ molecules.  Moreover, unusual configurations found in HMLP-MD with preliminary versions of the potential were added to the DFT training set to improve performance and transferability.

We emphasize that determining the suitable dataset is not straightforward.  Numerous iterations of the potential were required to obtain a good fit to the phase diagram.  A good fit to DFT forces of known phases is not evidence that other phases are unstable: rigorous testing of the potential in MD is essential.

\subsubsection*{Covalent bond}
The covalent bonding contribution to the force is approximated as $(F_1^{\mu_{12}}-F_2^{\mu_{12}})/2$, where  $F_1^{\mu_{12}}$ is the component of atomic force projected down the molecular axis. We examined various options for fitting the covalent bond: harmonic, Lennard-Jones and Morse potentials. The latter provides a better fit across our dataset using simple regression. 

\subsubsection*{Non-bonded interactions}
We describe the short-ranged Coulomb and van der Waals potentials using pairwise functions to create the fingerprint.  These are built using Gaussians with a smooth cutoff in the form
\begin{equation}
    V_i^k = \sum \exp(-|R_{ij}/\eta_k|^2)f_{cut}(R_{ij})\hat{\bf R}_{ij}
\end{equation}                         
where ${\bf R}_{ij}$ is the vector between molecules $i$ and $j$, with $\eta_k$ the range of the $k^{th}$ fingerprint.

These fingerprints are mapped onto the corresponding {\it residual} atomic forces,  defined by the DFT forces less the contribution from the Morse potentials.
This mapping is achieved using the Kernel Ridge Regression method which is capable of handling complex nonlinear relationships.  The set of $\eta_k$ is optimised using regularization and feature selection algorithms\cite{kira1992practical}.

\subsubsection*{Angle-dependent interactions}
Finally, we consider the orientation-dependent forces. These are fitted to the residuals once covalent and pairwise interactions are subtracted from the DFT forces. The corresponding fingerprints for the orientation-dependent interactions are again chosen using feature selection algorithms and Kernel Ridge regression.

\subsection{Machine Learned Molecular Dynamics:}
The  simulations were performed using periodic boundary conditions and a time step of 0.5 fs. The Nose-Hoover thermostat and the Parrinello-Rahman barostat\cite{parrinello1981polymorphic} were used for controlling temperature and pressure, respectively. All simulations were carried out using the LAMMPS package and the atomic configurations were visualised with the AtomEye package. Typical models of H$_2$ system was created with P21/c structure containing 72,576 molecules.To reproduce the entire temperature-pressure phase diagram, the NPT simulations of 1152-atom supercells of P21/c structure were carried out at selected temperatures and pressures, from which we can identify the corresponding stable phases and melting point via the Z-method\cite{belonoshko2006melting}. Furthermore, the phase-coexistence method\cite{mendelev2003development} with co-existing H$_2$ solid and liquid was adopted to determine the melting curve.

\subsection{Analysis of Molecular Dynamics:}

\begin{figure}[!htbp]
    \centering
    \includegraphics[width=0.8\textwidth]{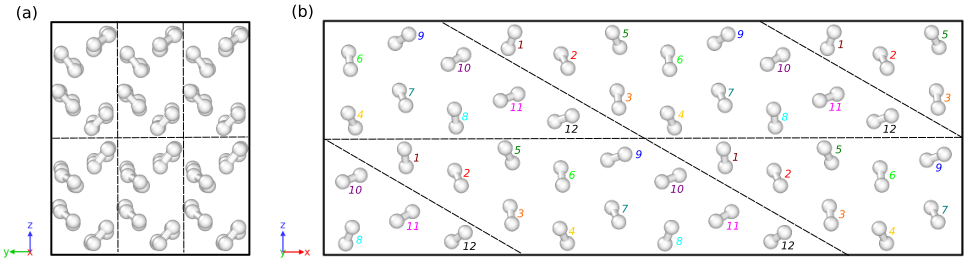}
    \caption{The 2x3x2 supercell of $P2_1/c\text{-}24$ used in this work shown in (a) top-down and (b) side views. This system contains twelve unit cells, denoted by the dashed lines. The twelve molecules that form the basis in each unit cell are numbered in the side view.}
    \label{fig:P21cbasis}
\end{figure}

To distinguish between the broken symmetry structure of phase II and the free rotors of phase I we introduce the orientational order parameter $O$. 
\begin{equation}
    \label{eq:OP}
    O=\left<\frac{\sum\limits_{b}\sum\limits_{i\neq j}(\hat{\mathbf{r}}_{i,b}\cdot\hat{\mathbf{r}}_{j,b})\mathbf{R}_{ij,b}}{\sum\limits_{b}\sum\limits_{i\neq j}\mathbf{R}_{ij,b}}\right>
\end{equation}
Here the summation is over unit cells $i,j$, each containing a set of basis molecules $b$. Unit vectors $\hat{\mathbf{r}}_{i,b}$ and $\hat{\mathbf{r}}_{j,b}$ are oriented along the the H-H bond of the $b^{th}$ molecule in the $i^{th}$ and $j^{th}$ unit cell, respectively, and $\mathbf{R}_{ij,b}$ is the distance between the center of mass of these two molecules. The angled brackets denote a time average. This parameter probes the long-range order in the system relative to the chosen basis, which in our case is the $P2_1/c\text{-}24$ unit cell.\cite{pickard2009structures} A value of 1 means that the system has the $P2_1/c\text{-}24$ structure, and a value of 0 suggests that the system is disordered. Note that this order parameter only detects similarity to the given basis and thus a phase change to a structure with a different unit cell will yield an erroneously low value. The trajectories were therefore visually inspected in addition to the order parameter analysis. A 2x3x2 supercell of $P2_1/c\text{-}24$ was used for phase II, and the unit cells and basis for this system are illustrated in Fig. \ref{fig:P21cbasis}.

 For the Phase I-II transition, MD trajectories were calculated for temperatures ranging from 10 - 150 K and pressures from 20 - 140 GPa. After a 5 ps equilibration period, the order parameter $O$ was averaged over the remaining 45 ps of the trajectory. The results are shown in Fig. \ref{fig:classicalOP_phaseDiagram}(a). 

The Raman-active phonons were extracted from the MD using the projection method which automatically includes anharmonic effects\cite{pinsook1999calculation,ackland2014efficacious}.

We tried numerous approaches to measure the orientation relationship between adjacent molecules, $i$ and $j$ with interatomic vectors $\vec{\sigma}_i$ and $\vec{\sigma}_j$ separated by $\hat{\vec{R}}$.  For Table \ref{table:EQQ} we used the explicit equation for linear quadrupoles: 
\begin{equation} \label{eq:eqq}
    E_{QQ} = \frac{3Q^2}{4\pi\epsilon_0}\sum_{i,j}\frac{\Gamma(\vec{\sigma_i},\vec{\sigma_j},\hat{\vec{R}})}{|\hat{\vec{R}}|^5}
\end{equation}{}
where $Q = 0.26 \text{D\AA}$ is the quadrupole moment of the $H_2$ molecule and the orientational factor $\Gamma(\vec{\sigma_i},\vec{\sigma_j},\hat{\vec{R}})$ is defined as:
\begin{equation} \label{eq:eqq-orient}
\begin{split}
\Gamma(\vec{\sigma}_i, \vec{\sigma}_j, \hat{\vec{R}}) &= 35 (\vec{\sigma}_i\cdot\hat{\vec{R}})^2(\vec{\sigma}_j\cdot\hat{\vec{R}})^2 -5(\vec{\sigma}_i\cdot\hat{\vec{R}})^2
-5(\vec{\sigma}_j\cdot\hat{\vec{R}})^2\\
&\,\,\,+2(\vec{\sigma}_i\cdot\vec{\sigma}_j)^2
-20(\vec{\sigma}_i\cdot\hat{\vec{R}})(\vec{\sigma}_j\cdot\hat{\vec{R}}) (\vec{\sigma}_i\cdot\vec{\sigma}_j) + 1
\end{split}
\end{equation}
While to estimate the steric hindrance, we calculated the average relative orientation of the $N$ hydrogen molecules to their 12 nearest neighbours as:
\begin{equation}
    X = \frac{1}{12N} \sum_{i,j}  \left[ 1 - \frac{1}{3} \Big( |\vec{\sigma_i}\cdot\vec{\sigma_j}| + |\vec{\sigma_i}\cdot\hat{\vec{R}}| + |\vec{\sigma_j}\cdot\hat{\vec{R}}| \Big) \right]
\end{equation}
X is scaled to equal 1 for a perfect "X" shape (i.e.  $\hat{\vec{\sigma}}_i = (1 0 0)$, $\hat{\vec{\sigma}}_i = (0 1 0)$ and $\hat{\vec{R_{ij}}} = (0 0 1)$\, ).  All other configurations have lower values for X, the least favoured being molecules pointing directly at one another (which has a value of 0).

\begin{addendum}
 \item  The authors acknowledge the ERC project HECATE for funding.  We are grateful for computational support from the UK national high performance computing service, ARCHER, and from the UK Materials and Molecular Modelling Hub, which is partially funded by EPSRC (EP/P020194), for both of which access was obtained via the UKCP consortium and funded by EPSRC grant ref EP/P022561/1.
\item[Competing Interests] The authors declare that they have no
competing financial interests.  We thank Miriam Pena-Alvarez and Eugene Gregoryanz for the experimental phonon and vibron data ahead of publication.  We thank Hua Geng for sharing complete DFT data from Ref.\cite{geng2019thermodynamic}.
 \item[Corresponding Author] Correspondence and requests for materials should be addressed H.Zong
\end{addendum}

\newpage
\bibliographystyle{naturemag}

\bibliography{Refs}

\section{Appendix: Phase III and other order parameters}
\begin{figure}[!htbp]
    \centering
    \includegraphics[width=\textwidth]{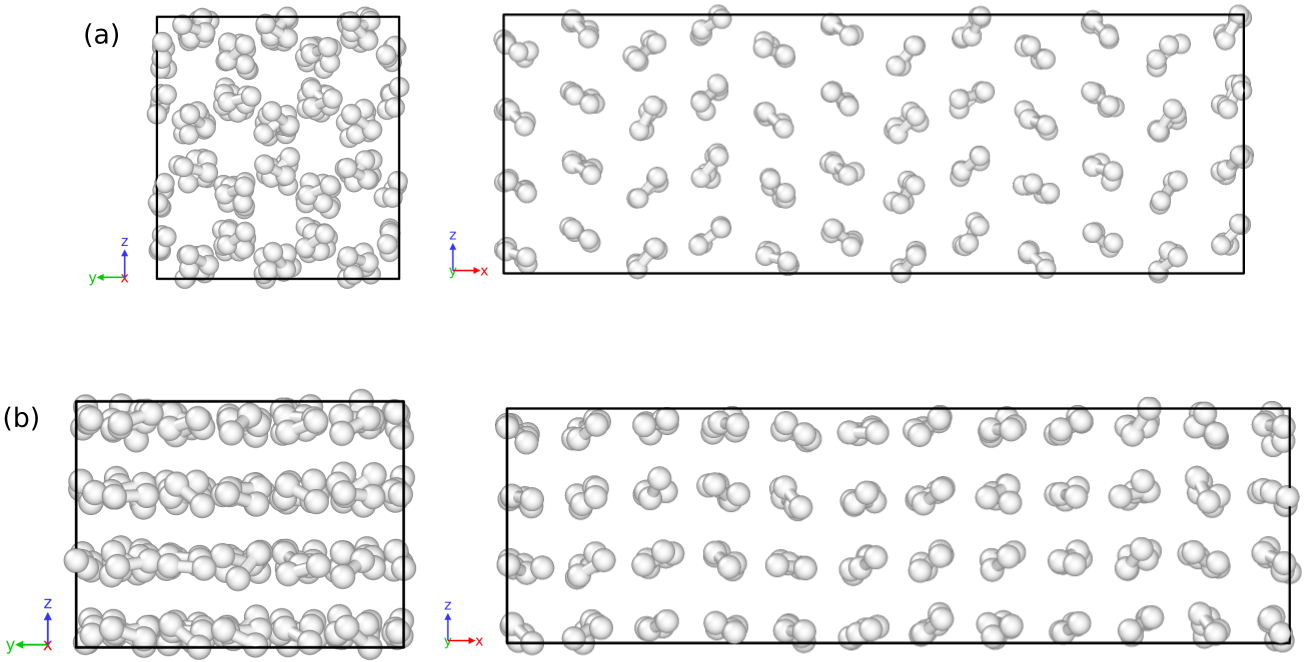}
    \caption{(Supplemental) Trajectory snapshots of the extra phase changes seen in the classical model. (a) The $P2_1/c$ structure seen for 60-100 GPa at intermediate temperatures. This example is taken from the 60 GPa, 60 K trajectory. (b) The "phase III-like" structure seen at high pressure and high temperature. This example is taken from the 120 GPa, 120 K trajectory.}
    \label{fig:classicalOtherPhases}
\end{figure}

Included here are some supplemental figures showing the phase transition in closer detail for the $O$ parameter and also for fluctuations in the system volume: $\langle V^2 \rangle - \langle V \rangle^2$. The phase lines are drawn based on visualizations of the trajectory as it is quite clear to the eye which phase the system is in. In most cases the boundary corresponds to either jumps in $O$ or to the peak in $\langle V^2 \rangle - \langle V \rangle^2$. There are two extra phase changes occurring which complicate interpretation of these plots. First is the transition to a $P2_1/c$ structure with a smaller unit cell (mentioned in the main text) for 60, 80 and 100 GPa at intermediate temperatures. However, for high pressures (120 and 140 GPa) and high temperatures (100-130 K) the system collapses to a close packed phase III-like structure as shown in Fig. \ref{fig:classicalOtherPhases}(b). 

\begin{figure}[!htbp]
    \centering
    \includegraphics[width=0.75\textwidth]{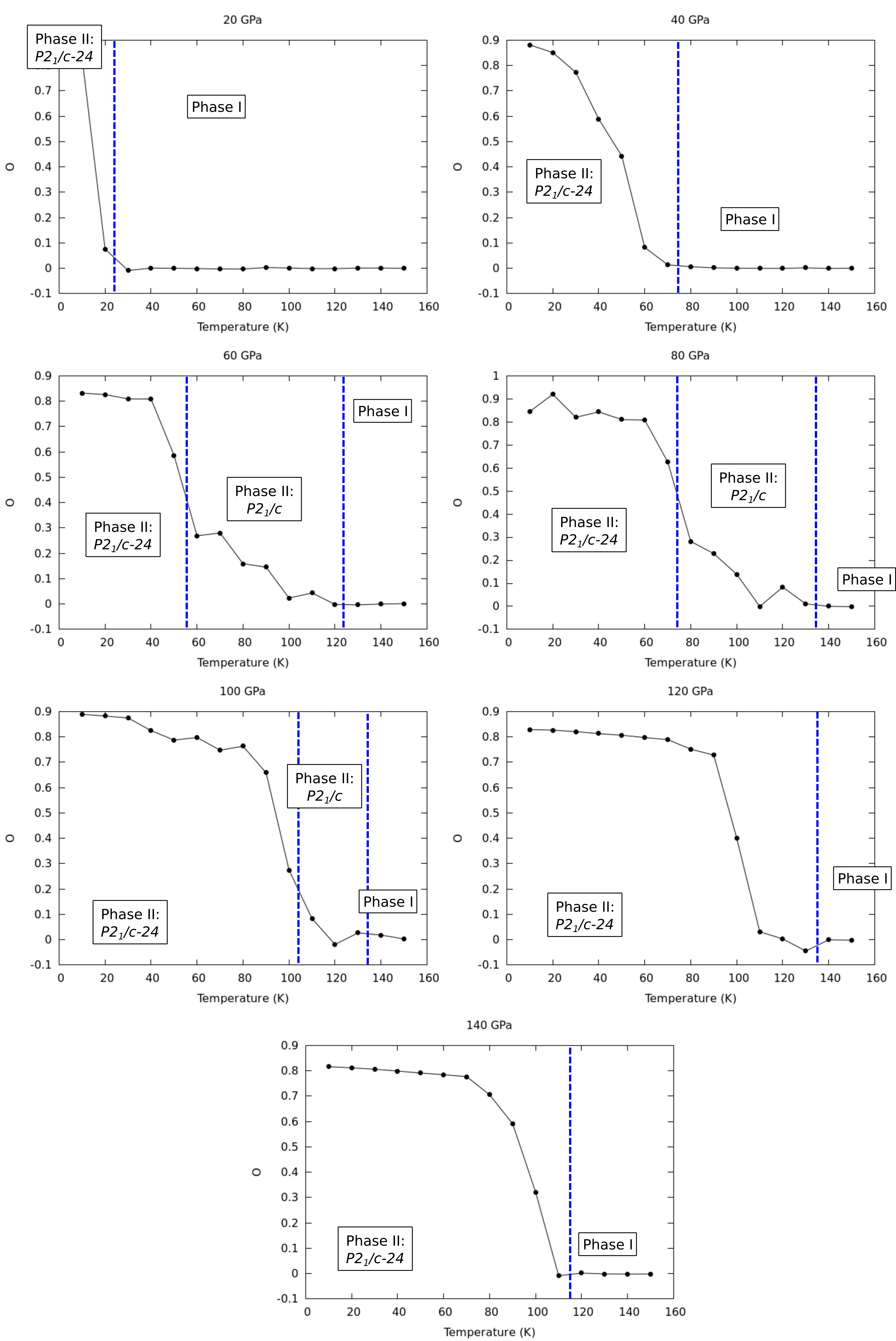}
    \caption{(Supplemental) Individual plots of $O$ for each pressure in the classical system. Phase lines are drawn based on a visualization of the trajectory as discontinuous jumps in $O$ for the higher pressure cases are due to phase changes within phase II and not a change to phase I.}.
    \label{fig:classicalOP}
\end{figure}

\begin{figure}[!htbp]
    \includegraphics[width=0.75\textwidth]{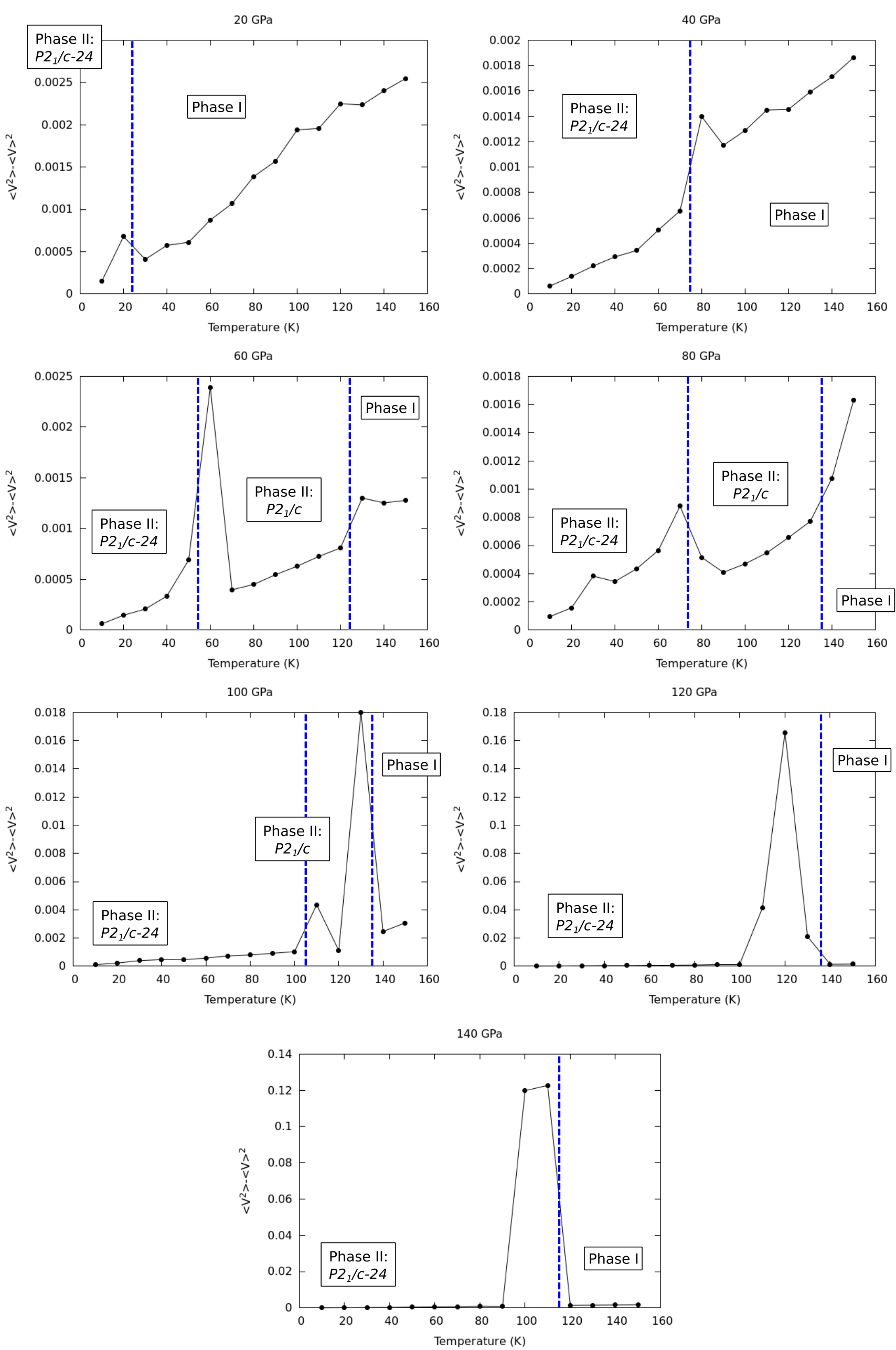}
    \caption{ (Supplemental) The phase transition in the classical system was also detected by a peak in the fluctuations of the volume $\langle V^2 \rangle - \langle V \rangle ^2$. The presence of two peaks in the 60, 80 and 100 GPa plots indicate the transition to a different $P2_1/c$ structure. The very large peaks in the 100, 120 and 140 GPa plots are the transition to the "phase III" close packed structure.}.
    \label{fig:classicalVfluct}
\end{figure}

\end{document}